\documentclass[%
 reprint,
 superscriptaddress,
 amsmath,amssymb,
 aps,
]{revtex4-1}

\usepackage{graphicx}% Include figure files
\usepackage{dcolumn}% Align table columns on decimal point
\usepackage{bm}% bold math
\usepackage{xcolor}
\usepackage{url}
\usepackage{subfigure}
\usepackage{multirow}

\begin{document}

%\preprint{APS/123-QED}

\title{Proton and deuteron mass radii from near-threshold $\phi$-meson photoproduction}%% Force line breaks with \\
%%\thanks{A footnote to the article title}%

\author{Rong Wang}
\email{rwang@impcas.ac.cn}
\affiliation{Institute of Modern Physics, Chinese Academy of Sciences, Lanzhou 730000, China}
\affiliation{University of Chinese Academy of Sciences, Beijing 100049, China}

\author{Wei Kou}
\email{kouwei@impcas.ac.cn}
\affiliation{Institute of Modern Physics, Chinese Academy of Sciences, Lanzhou 730000, China}
\affiliation{University of Chinese Academy of Sciences, Beijing 100049, China}

\author{Chengdong Han}
\email{chdhan@impcas.ac.cn}
\affiliation{Institute of Modern Physics, Chinese Academy of Sciences, Lanzhou 730000, China}
\affiliation{University of Chinese Academy of Sciences, Beijing 100049, China}

\author{Jarah Evslin}
\email{jarah@impcas.ac.cn}
\affiliation{Institute of Modern Physics, Chinese Academy of Sciences, Lanzhou 730000, China}
\affiliation{University of Chinese Academy of Sciences, Beijing 100049, China}

\author{Xurong Chen}
\email{xchen@impcas.ac.cn}
\affiliation{Institute of Modern Physics, Chinese Academy of Sciences, Lanzhou 730000, China}
\affiliation{University of Chinese Academy of Sciences, Beijing 100049, China}
\affiliation{Guangdong Provincial Key Laboratory of Nuclear Science, Institute of Quantum Matter, South China Normal University, Guangzhou 510006, China}


\date{\today}% It is always \today, today,

\begin{abstract}
\noindent
We analyze the exclusive $\phi$-meson photoproduction on both hydrogen and deuterium targets
based on the published data of CLAS, SAPHIR and LEPS collaborations.
A dipole-form scalar gravitational form factor is applied to describe
the $|t|$-dependence of the differential cross section.
Based on the average of all the near-threshold data from CLAS, SAPHIR and LEPS,
we find that the proton and deuteron mass radii are $0.75 \pm 0.02$ fm
and $1.95 \pm 0.19$ fm respectively.
The coherent and near-threshold quarkonium photoproduction seems
to be sensitive to the radius of the hadronic system.
The vector-meson-dominance model together with a low-energy QCD theorem
well describe the data of near-threshold $\phi$
photoproduction on the hadronic systems.
\end{abstract}

\pacs{12.38.-t, 14.20.Dh}% PACS, the Physics and Astronomy
%% \keywords{Suggested keywords}%Use showkeys class option if keyword
\maketitle

%\tableofcontents

Hadronic matter accounts for nearly all the mass of the visible universe.
However many mysteries remain concerning hadron structure, such as
how the hadron mass, spin and pressure are fixed by strong interactions
of the quarks and gluons inside
\cite{Ji:1994av,Ji:1995sv,Ji:2021pys,Ji:1996nm,Ji:1997pf,Ji:1996ek,Teryaev:2016edw,Polyakov:2002yz,Polyakov:2018zvc,Burkert:2018bqq}.
The inside of a hadron is governed by quantum chromodynamics (QCD) theory.
In particular, the strong force carrier in QCD, the gluon, is believed to play
a dominant role in determining the basic properties of the proton, such as its mass.
X. Ji found that the proton mass scale is fixed by the QCD quantum trace anomaly
\cite{Ji:1994av,Ji:1995sv,Ji:2021pys,Yang:2018nqn,Wang:2019mza,Rodini:2020pis,Metz:2020vxd}.

The mechanical properties of the hadronic system are well encoded
in the energy-momentum tensor (EMT) $T^{\mu\nu}$,
and the gravitational form factors (GFF) are the EMT matrix elements
in the hadron state $\left<p^{\prime}|T^{\mu\nu}|p\right>$
\cite{Kobzarev:1962wt,Pagels:1966zza,Ji:1996nm,Teryaev:2016edw,Polyakov:2002yz,Polyakov:2018zvc}.
The EMT characterizes the response of the hadronic system to the change of the space-time metric,
hence the GFF could be directly accessed via the graviton scattering
but with almost zero hope in reality.
GFF contains three fundamental properties of the hadron:
mass, spin and D-term (pressure and shear)
\cite{Pagels:1966zza,Ji:1996nm,Ji:1997pf,Teryaev:2016edw,Polyakov:2002yz,Polyakov:2018zvc}.
Due to the significance of GFF, searching for the experimental ways of constraining
GFF is of high interests.

Now the proton mass radius is the subject of heated discussions.
This mass radius characterizes how the mass is distributed inside the proton,
namely the mass density distribution.
Experimentally, the form factors as a function of momentum transfer ($q$)
are measured in elastic scattering processes, which correspond to
the Fourier transforms of the various density distributions.
Theoretically, the mass radius of a hadron can be defined in terms of
the scalar GFF $G(t=q^2)$, the form factor of the trace
of the QCD EMT, in a nonrelativistic and weak gravitational field approximation \cite{Kharzeev:2021qkd}.
Similar to the definition of the charge radius, the mass radius is defined as
the derivative of the form factor at zero momentum transfer \cite{Kharzeev:2021qkd,Miller:2018ybm},
\begin{equation}
\begin{split}
\left<R_{\rm m}^2\right> \equiv \frac{6}{M}\frac{dG(t)}{dt}\big|_{t=0},
\end{split}
\label{eq:MassRadius}
\end{equation}
with $G(0)=M$.
Theoretically, the density distribution inside the proton is the expectation value
of the component $T^{00}$ of EMT.
Assuming local equilibrium, the EMT can be expanded in a gradient expansion,
and it is entirely characterized by the density and the pressure at leading order of the expansion.
For the stable system, the pressure-volume work vanishes.
In this case, in the Landau frame, $T^{00}$ is equal to the trace $T^{\mu}_{\mu}$ of EMT.
Therefore it is not surprising that, at small momentum transfer, the density distribution
may be constrained by an experimental determination of the scalar form factor $\left<p^{\prime}|T^{\mu}_{\mu}|p\right>$.

Following the proposal by D. Kharzeev \cite{Kharzeev:2021qkd,Kharzeev:1995ij,Kharzeev:1998bz,Fujii:1999xn},
we describe the near-threshold cross section for heavy quarkonium
photoproduction using the QCD Van der Waals force of leading-twist gluon operators
derived from a QCD low energy theorem and a vector-meson-dominance (VMD) model.
The theoretical uncertainties are argued to be under control in the nonrelativistic limit.
The differential cross section for quarkonium photoproduction
in the small-$|t|$ region and near the production threshold can be described with
the scalar GFF $G(t)$ \cite{Kharzeev:2021qkd}, which is written as,
\begin{equation}
\begin{split}
\frac{d\sigma}{dt} \propto G^{2}(t).
\end{split}
\label{eq:DiffXsection}
\end{equation}
Therefore the near-threshold quarkonium photoproduction data
is sensitive to the GFF and the mass distribution.
With this theoretical framework and the dipole form of GFF, $G(t)=M/(1-t/m_{\rm s}^2)^2$,
the mass radius of the proton has been extracted from
vector meson production data on the hydrogen targets \cite{Kharzeev:2021qkd,Wang:2021dis}.
By analyzing the GlueX data of J/$\psi$ photoproduction near the threshold \cite{GlueX:2019mkq},
Kharzeev extracted the mass radius of the proton to be $0.55\pm 0.03$ fm \cite{Kharzeev:2021qkd}.

The VMD model is successful in describing the vector meson photoproduction process,
at least as the first step of more extensive theoretical studies.
In the VMD model, a real photon fluctuates into a virtual vector meson,
which subsequently scatters off the proton target.
This model assumption is based on the fact that the vector meson
has the same quantum numbers of the photon.
Recently, the VMD model was used for determining the $\omega-p$, $\phi-p$
and J/$\psi-p$ scattering lengths \cite{Strakovsky:2014wja,Strakovsky:2020uqs,Strakovsky:2019bev}.
These analyses show that the J/$\psi-p$ scattering length is much smaller than
that of $\omega-p$ and $\phi-p$. This suppressed interaction in $c\bar{c}-p$ system
can be explained with the ``young'' vector meson effect \cite{Strakovsky:2021vyk,Feinberg:1980yu}.
If the mass of the quark pair is far from the photon mass,
then the vector meson - proton interaction is strongly suppressed
due to the small size of the ``young'' vector meson just created by the photon.
The $c\bar{c}$ pair lacks sufficient time to form the complete wave function of J/$\psi$,
and its radius is much smaller than that of J/$\psi$.
To avoid the influence that may associated with the ``young'' vector meson effect,
it is necessary and worthy trying to look at the light vector meson photoproduction.

The $\phi$ vector meson is a typical quarkonium which do not has
the same type of quark of the proton valence components.
Thus the $\phi$ meson interacts with the proton via the gluon exchanges
which is argued to be sensitive to the scalar GFF of the proton.
Moreover, the minimum momentum transfer square $|t|$ is around 0.2 GeV$^2$,
for the current available $\phi$-photoproduction data near threshold.
The large $|t|$ makes the use of VMD model questionable.
Hence the data of small $|t|$ is better for the study of the interaction
between the vector meson and the proton, applying the VMD model.

An interesting question is whether or not the scalar GFF is also applicable
to the near-threshold quarkonium production on the atomic nucleus (such as the deuteron).
The CLAS and LEPS collaborations have measured the coherent
$\phi$-photoproduction cross sections on hydrogen and deuterium targets
\cite{CLAS:2013jlg,CLAS:2007xhu,LEPS:2005hax,Chang:2007fc}.
Due to the high luminosity of the accelerator and the large acceptance
of the CLAS spectrometer, the CLAS data is of high precision
over a wide $|t|$ range \cite{CLAS:2013jlg,CLAS:2007xhu}.
SAPHIR collaboration also has measured the $\phi$-photoproduction
on the hydrogen target near the reaction threshold \cite{Barth:2003bq}.
These exclusive $\phi$-photoproduction data provide an excellent opportunity
to determine the mass radii of the proton and the deuteron,
and to examine the scalar GFF interpretation of
the quarkonium production mechanism in the nonperturbative region.

In this paper, we present simultaneous analyses of the proton mass radius
and the deuteron mass radius using CLAS, SAPHIR and LEPS data,
following the previous studies on the proton mass radius \cite{Kharzeev:2021qkd,Wang:2021dis}.
The systematic uncertainty from varying the $|t|$ range of the fit is also investigated,
thanks to the wide $t$ range of CLAS data.

\begin{figure}[htp]
\centering
\includegraphics[width=0.46\textwidth]{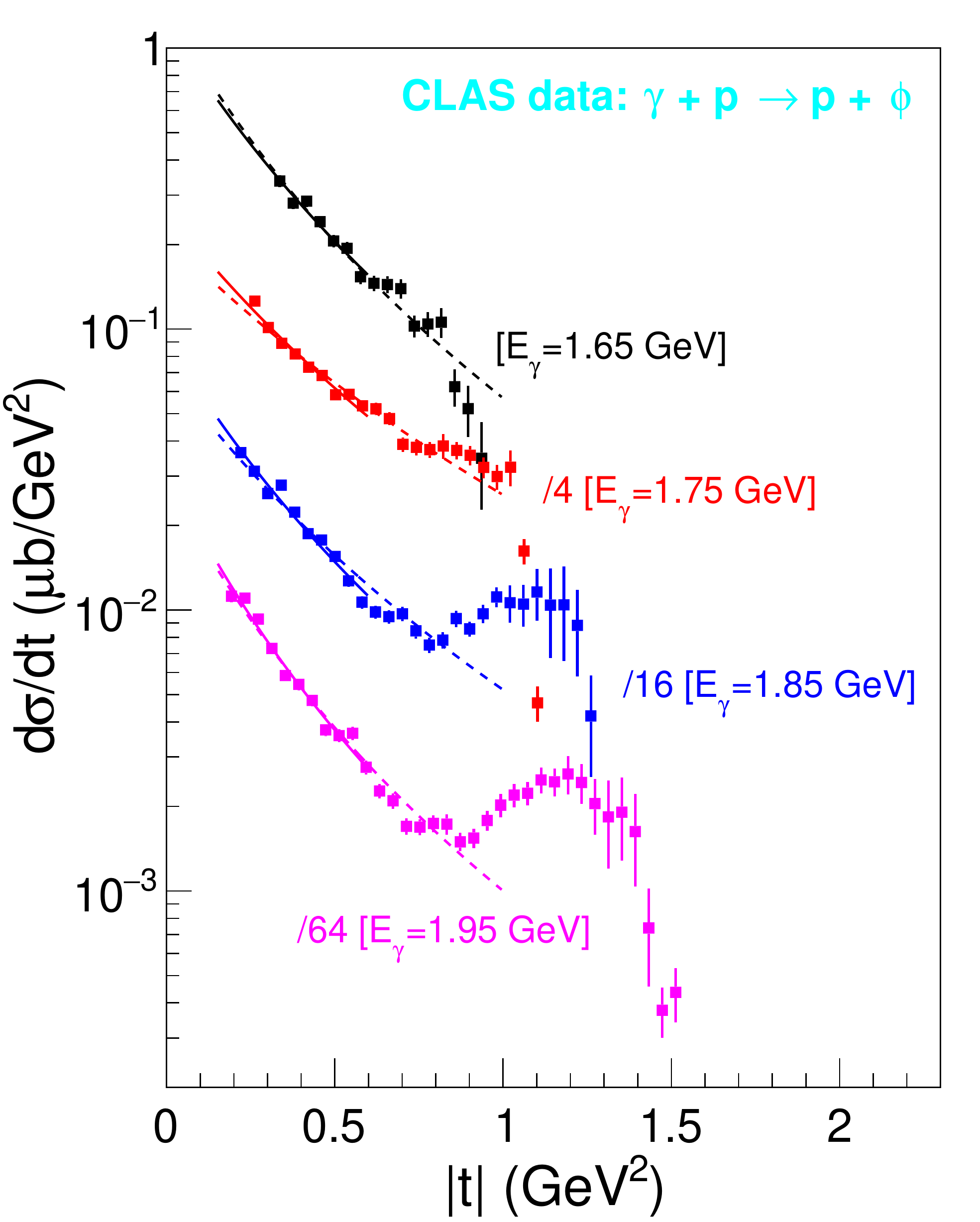}
\caption{
The differential cross sections of near-threshold $\phi$ photoproduction
on a hydrogen target ($\gamma p \rightarrow \phi p$) from CLAS \cite{CLAS:2013jlg}.
Only the statistical errors are presented.
Some of the cross sections are scaled by factors indicated in the figure.
The dashed curves show the fits in the $|t|$ range from 0.15 GeV$^2$ to 1 GeV$^2$.
The solid curves show the fits in the $|t|$ range from 0.15 GeV$^2$ to 0.6 GeV$^2$.
}
\label{fig:diff-xsection-proton-target-CLAS}
\end{figure}

\begin{figure}[htp]
\centering
\includegraphics[width=0.42\textwidth]{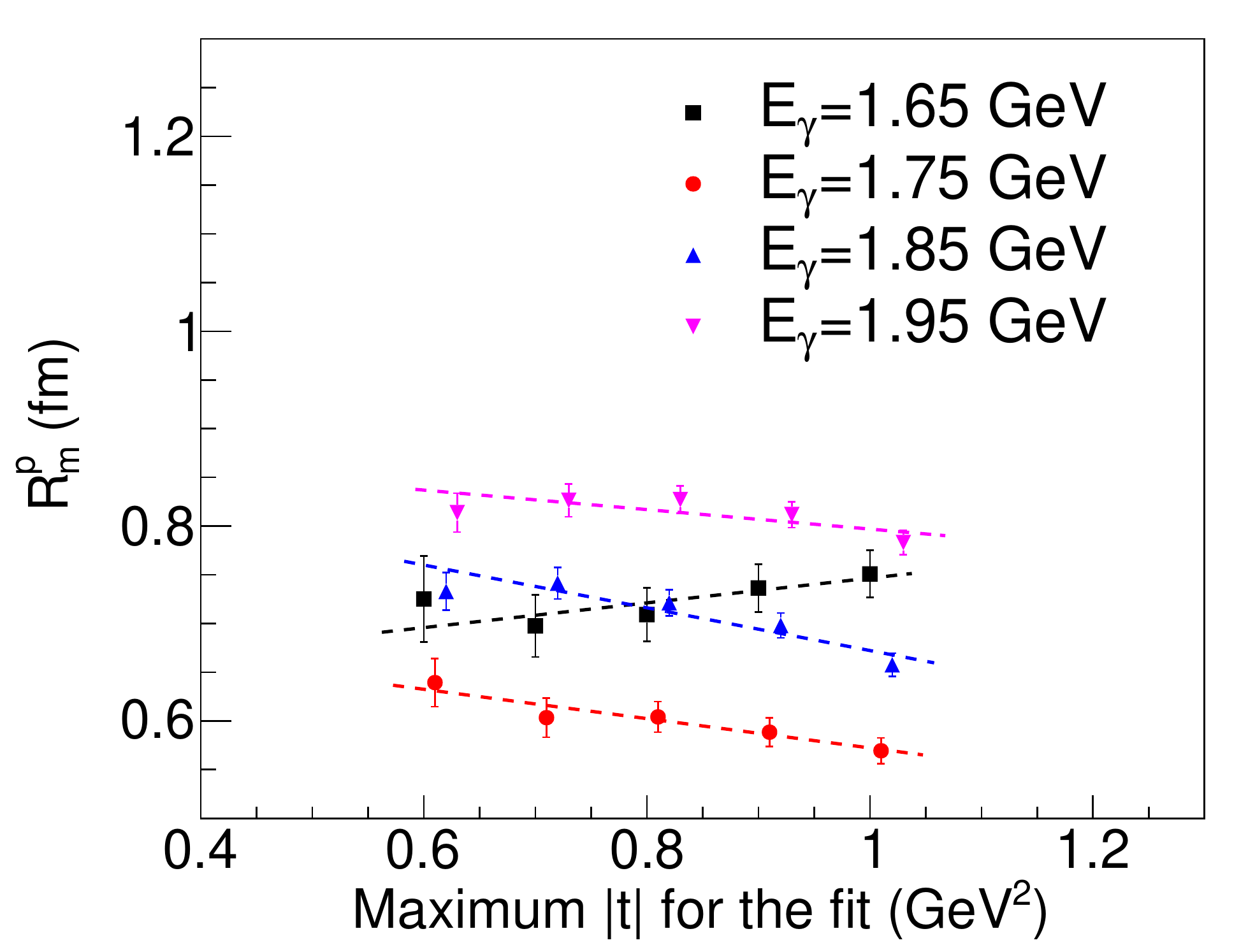}
\caption{
The proton mass radius as a function of the maximum $|t|$ in the fitting range,
extracted from CLAS data.
To avoid overlaps of the data points,
the data points at 1.75 GeV, 1.85 GeV, and 1.95 GeV are shifted
to the right by 0.01 GeV$^2$, 0.02 GeV$^2$ and 0.03 GeV$^2$ respectively.
The dashed lines display the linear fits of the data.
}
\label{fig:proton-mass-radius-and-t-range}
\end{figure}

\begin{table}[h]
\caption{The proton mass radii at different photon energies
determined from the $t$-slopes of the differential cross sections from CLAS,
extrapolated to $t=0$ GeV$^2$.
}
\begin{center}
\begin{ruledtabular}
\begin{tabular}{ ccccc }
  $E_{\gamma}$ (GeV)  &    1.65         &    1.75        &      1.85         &       1.95     \\
  \hline
  $R_{\rm m}^{\rm p}$ (fm)    &  $0.62\pm0.09$  &  $0.72\pm0.05$ &    $0.89\pm0.04$  &   $0.89\pm0.05$    \\
\end{tabular}
\end{ruledtabular}
\end{center}
\label{tab:ProtonMassRadius-CLAS}
\end{table}

Fig. \ref{fig:diff-xsection-proton-target-CLAS} shows the CLAS collaboration's near-threshold
differential cross section data for $\phi$ photoproduction on the proton.
It has been suggested that the $|t|$-dependence of the differential cross section is
described by the scalar GFF of the proton.
Some fits based on Eq. (\ref{eq:DiffXsection}) are shown in the figure.
The differential cross section data are well reproduced
by the dipole form of the scalar GFF.
However in the large-$|t|$ region, approaching $|t|_{\rm max.}$,
we see the cross section rising with $|t|$.
This behavior in the large-$|t|$ region may be due to the direct $\phi$-radiation contributions
from $u$-channel and $s$-channel with $\phi N N$ coupling or $\phi N N^*$ coupling
\cite{CLAS:2000kid,Laget:2000gj,Zhao:2001ue,Titov:2003bk,Williams:1998ge,Titov:1998tx,Titov:1999eu,Oh:2001bq}.

In order to avoid such $u$-channel or $s$-channel contamination,
we narrow the fitting range of $|t|$ towards the small-$|t|$ region.
We studied the fits restricted in quite different $|t|$ ranges carefully,
as the CLAS data covers a wide $|t|$ range and of high precision.
To understand the effect of the large-$|t|$ data on
the extraction of the scalar GFF,
we perform a series of fits excluding the large-$|t|$ data requiring
$|t|<0.6$ GeV$^2$, $|t|<0.7$ GeV$^2$, $|t|<0.8$ GeV$^2$, $|t|<0.9$ GeV$^2$,
and $|t|<1.0$ GeV$^2$ respectively.
With each fit, we extract the parameterized scalar GFF and
by calculating the derivative at $t=0$ GeV$^2$ we find the mass radius.
The obtained proton mass radii from these fits are shown
in Fig. \ref{fig:proton-mass-radius-and-t-range}, as a function of the cut at large $|t|$.
We find that, beneath 1 GeV$^2$, the extracted mass radius does not depend
strongly on the choice of large-$|t|$ cut.
This is probably because the $u$-channel or $s$-channel contribution only dominates
in the large-$|t|$ region where the error bars are comparatively large
\cite{Laget:2000gj,Zhao:2001ue,Titov:2003bk,Williams:1998ge,Titov:1998tx,Titov:1999eu,Oh:2001bq}.
Therefore the large-$|t|$ data hardly affects the fits.
The uncertainty from varying the fitting range of $|t|$ is
of the same order as the statistical uncertainty in the data.

With the linear extrapolations of the dependence on the fitting range shown in Fig. \ref{fig:proton-mass-radius-and-t-range},
we provide the mass radii in Table \ref{tab:ProtonMassRadius-CLAS}.
We see that the extracted mass radius depends weakly on the energy of the incident photon.
Since the formulism based on the GFF works mainly in the low energy limit,
the reliable mass radius should be given by the data closest to the production threshold.
Hence, from the CLAS data, the mass radius of the proton should be $0.62\pm0.09$ fm,
as determined from the $\phi$ photoproduction data at $E_{\gamma}=1.65$ GeV
(the most near-threshold photon energy of CLAS data).

\begin{figure}[htp]
\centering
\includegraphics[width=0.46\textwidth]{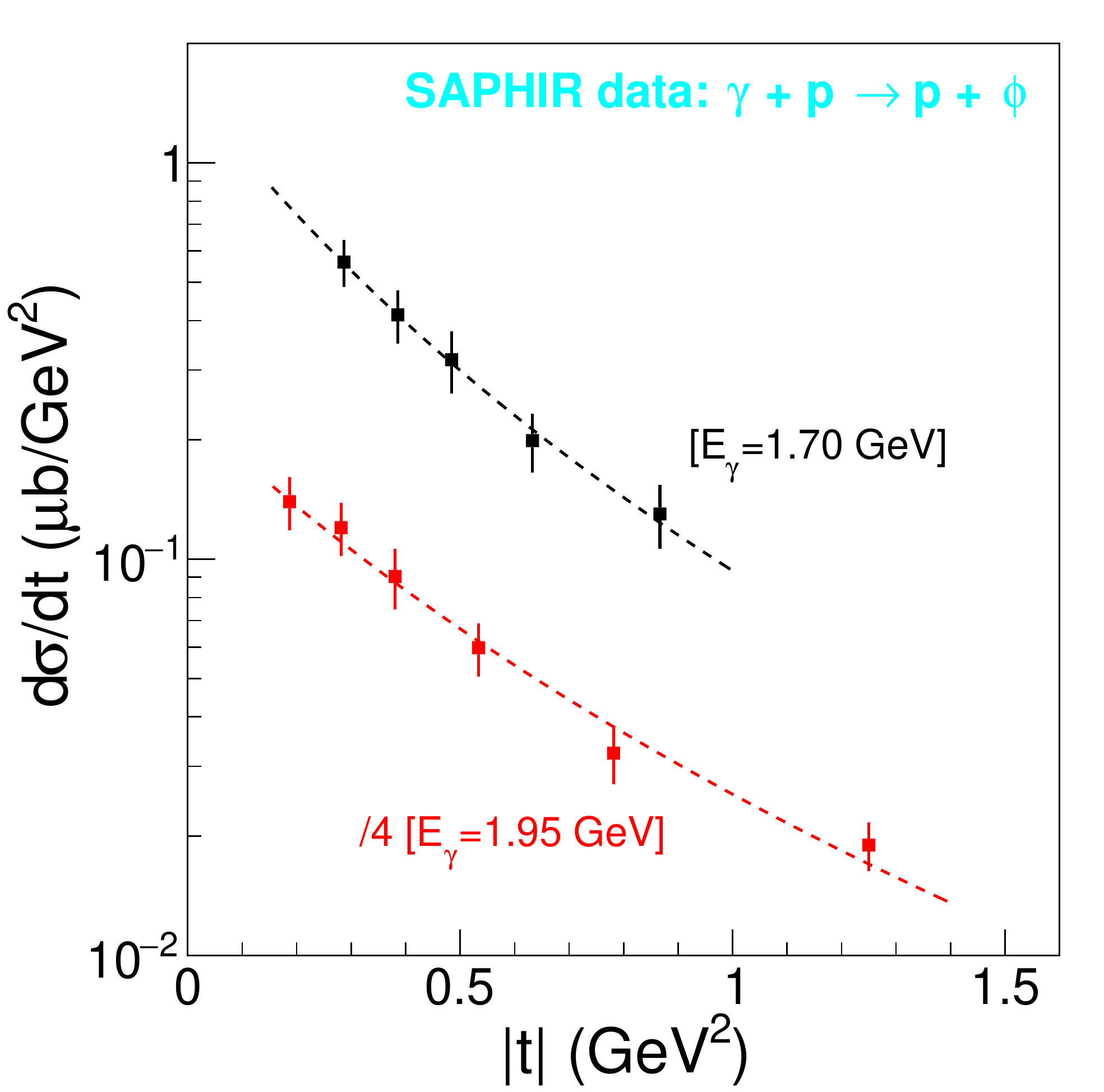}
\caption{
The differential cross sections of exclusive $\phi$ photoproduction
on proton near threshold ($\gamma p \rightarrow \phi p$) from SAPHIR \cite{Barth:2003bq}.
Only the statistical errors are presented.
The cross-section data at $E_{\gamma}=1.95$ GeV is divided by 4,
as indicated in the figure.
The dashed curves show the fits of scalar GFF.
}
\label{fig:diff-xsection-deuteron-target-SAPHIR}
\end{figure}

The proton mass radius from the proton data of LEPS collaboration is provided
in the previous analysis, which is $0.67 \pm 0.10$ in average \cite{Wang:2021dis}.
Though with fewer data points and larger statistical uncertainties,
SAPHIR collaboration has also measured the near-threshold $\phi$ photoproduction
on the proton two decades ago \cite{Barth:2003bq}.
The differential cross section data of SAPHIR are shown
in Fig. \ref{fig:diff-xsection-deuteron-target-SAPHIR}.
From the fits of dipole GFF, the SAPHIR data give the proton mass radii
to be $0.69\pm 0.08$ fm and $0.59\pm 0.04$ fm at $E_{\gamma}=1.7$ GeV and $E_{\gamma}=1.95$ GeV, respectively
(summarized in Table \ref{tab:ProtonMassRadius-SAPHIR}).
The CLAS, LEPS and SAPHIR data are more or less consistent with each other.

By looking at the average of all the near-threshold data of CLAS, LEPS and SAPHIR,
the proton mass radius is combined to be $0.75 \pm 0.02$ fm.
Compared to the analysis of the J/$\psi$-photoproduction data \cite{Kharzeev:2021qkd},
the extracted proton mass radius from $\phi$-photoproduction data
is much smaller. This discrepancy is maybe due to the different strengths of
the ``young'' vector meson effect for J/$\psi$ and $\phi$,
or that the VMD model is no longer effective in the large $|t|$ region.
More theoretical studies are needed to understand the dependence on
the size of the vector meson probe and the validity of applying VMD model in case of J/$\psi$.

\begin{table}[h]
\caption{The proton mass radii at different photon energies
determined from the $t$-slopes of the differential cross sections from SAPHIR.
}
\begin{center}
\begin{ruledtabular}
\begin{tabular}{ ccc }
  $E_{\gamma}$ (GeV)          &  $\left[1.57,1.82\right]$  & $\left[1.82,2.07\right]$   \\
  \hline
  $R_{\rm m}^{\rm p}$ (fm)    &      $0.69\pm 0.08$        &     $0.59\pm 0.04$        \\
\end{tabular}
\end{ruledtabular}
\end{center}
\label{tab:ProtonMassRadius-SAPHIR}
\end{table}

\begin{figure}[htp]
\centering
\includegraphics[width=0.46\textwidth]{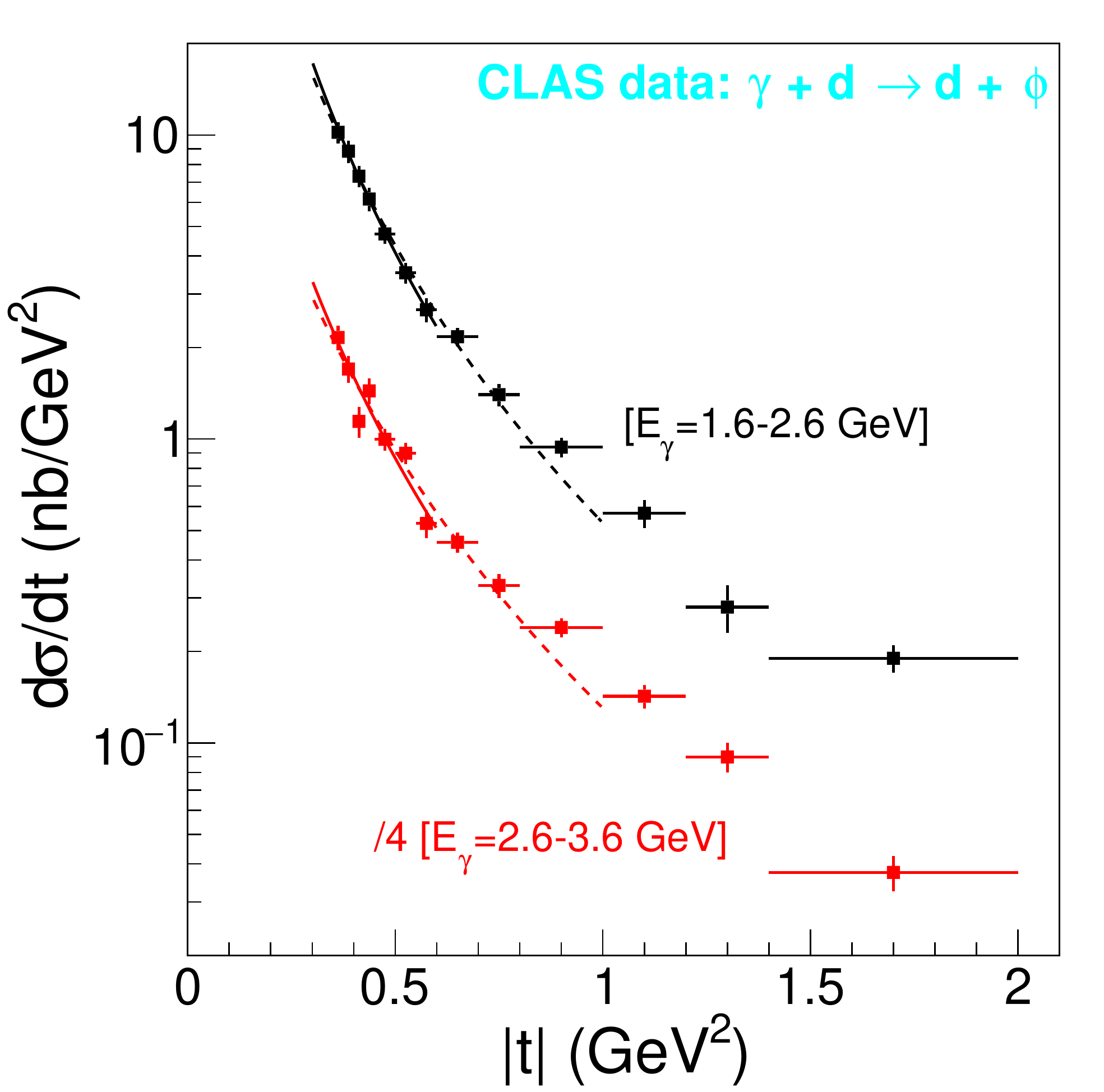}
\caption{
The differential cross sections of coherent $\phi$ photoproduction
on deuteron near threshold ($\gamma d \rightarrow \phi d$) from CLAS \cite{CLAS:2007xhu}.
Only the statistical errors are presented.
The cross-section data at $E_{\gamma}=3.1$ GeV is divided by 4,
as indicated in the figure.
The dashed curves show the fits in the $|t|$ range from 0.3 GeV$^2$ to 1 GeV$^2$.
The solid curves show the fits in the $|t|$ range from 0.3 GeV$^2$ to 0.6 GeV$^2$.
}
\label{fig:diff-xsection-deuteron-target-CLAS}
\end{figure}

\begin{figure}[htp]
\centering
\includegraphics[width=0.42\textwidth]{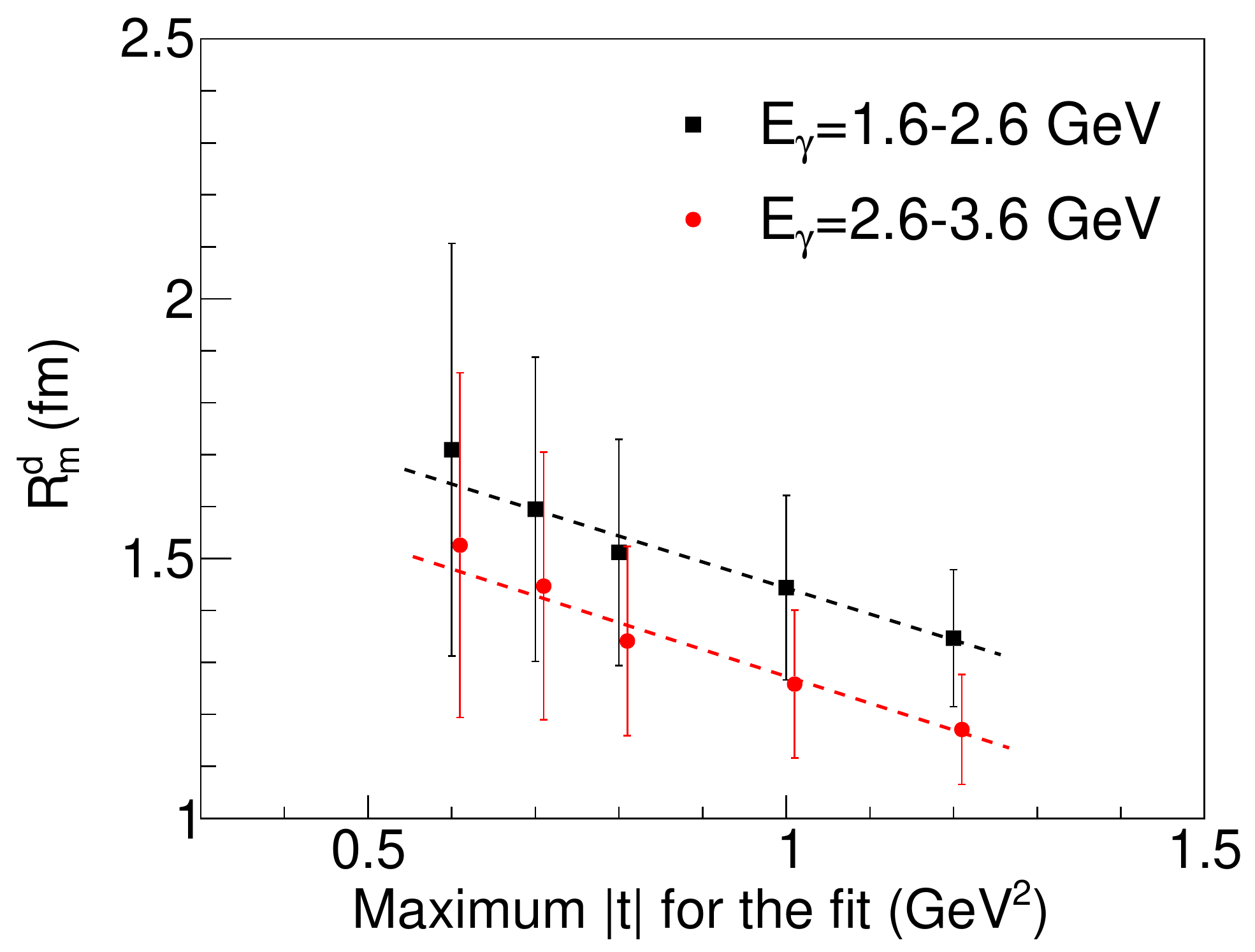}
\caption{
The deuteron mass radius as a function of the maximum $|t|$ in the fitting range,
extracted from CLAS data.
To avoid overlaps of the data points,
the data points of the energy bin [2.6, 3.6] GeV are shifted
to the right by 0.01 GeV$^2$.
The dashed lines display the linear fits of the data.
}
\label{fig:deuteron-mass-radius-and-t-range}
\end{figure}

\begin{table}[h]
\caption{The deuteron mass radii at different photon energies
determined from the $t$-slopes of the differential cross sections from CLAS,
extrapolated to $t=0$ GeV$^2$.
}
\begin{center}
\begin{ruledtabular}
\begin{tabular}{ ccc }
  $E_{\gamma}$ (GeV)  &    $\left[1.6,2.6\right]$   &    $\left[2.6,3.6\right]$     \\
  \hline
  $R_{\rm m}^{\rm d}$ (fm)    &  $1.94\pm0.45$  &  $1.78\pm0.38$   \\
\end{tabular}
\end{ruledtabular}
\end{center}
\label{tab:DeuteronMassRadius-CLAS}
\end{table}

\begin{figure}[htp]
\centering
\includegraphics[width=0.46\textwidth]{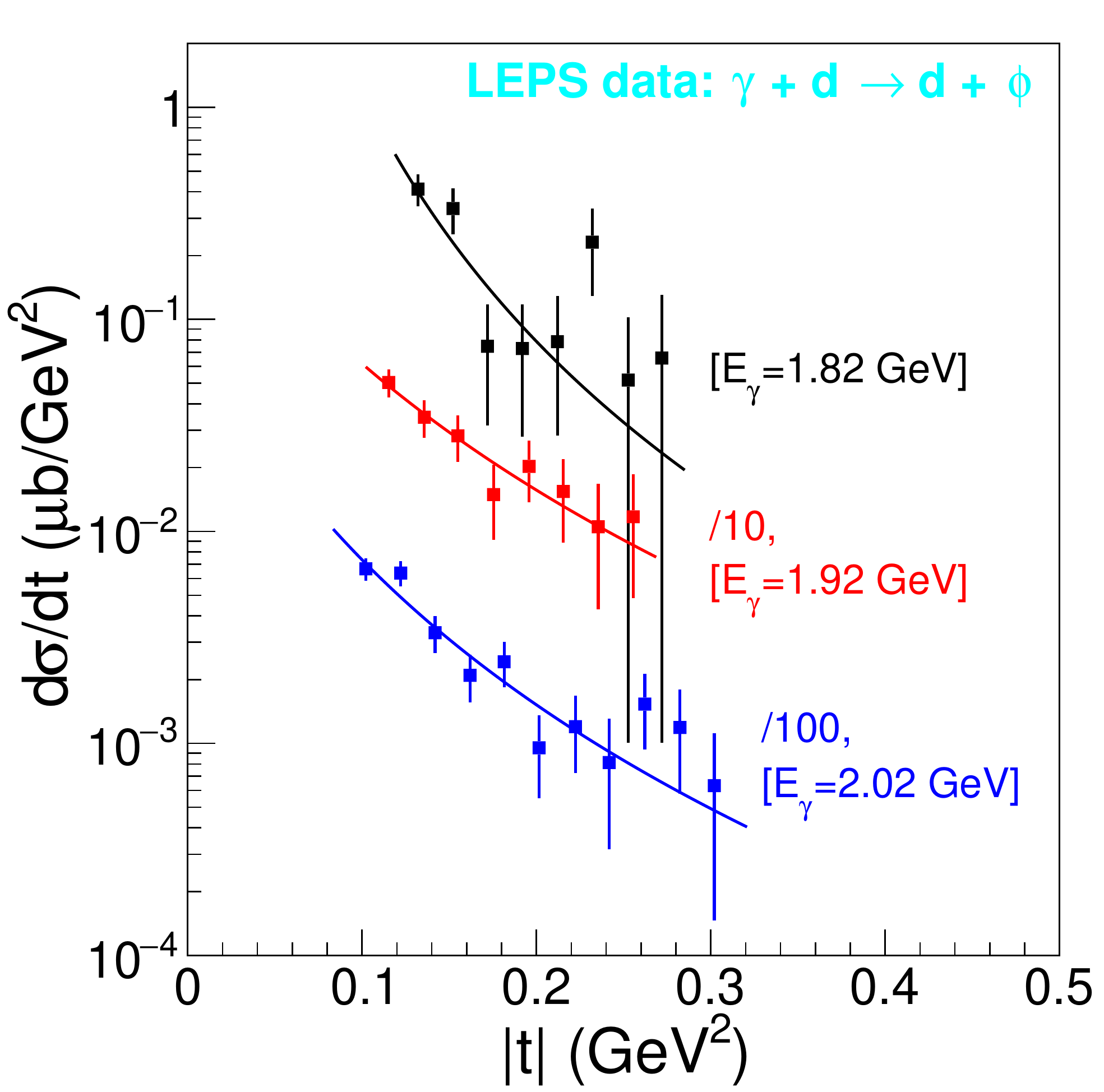}
\caption{
The differential cross sections of coherent $\phi$ photoproduction
on deuteron near threshold ($\gamma d \rightarrow \phi d$) from LEPS \cite{Chang:2007fc}.
Only the statistical errors are presented.
The cross-section data at $E_{\gamma}=1.92$ GeV and $E_{\gamma}=2.02$ GeV
are divided by 10 and 100 respectively,
as indicated in the figure.
The curves show the fits of scalar GFF.
}
\label{fig:diff-xsection-deuteron-target-LEPS}
\end{figure}

\begin{table}[h]
\caption{The deuteron mass radii at different photon energies
determined from the $t$-slopes of the differential cross sections from LEPS.
}
\begin{center}
\begin{ruledtabular}
\begin{tabular}{ cccc }
  $E_{\gamma}$ (GeV)          &       $1.82$   &    $1.92$      &  $2.02$           \\
  \hline
  $R_{\rm m}^{\rm d}$ (fm)    &  $11.7\pm4.2$  &  $1.80\pm0.43$ &    $2.08\pm0.30$     \\
\end{tabular}
\end{ruledtabular}
\end{center}
\label{tab:DeuteronMassRadius-LEPS}
\end{table}

In the following, we perform a similar analysis on the deuteron data.
First, we fit the differential cross section data on the deuteron
of CLAS collaboration, applying the VMD model and the scalar GFF.
Still surprisingly, the dipole form GFF describes well the coherent $\phi$
photoproduction data on the deuterium target, which is shown in Fig. \ref{fig:diff-xsection-deuteron-target-CLAS}.
Second, we study the mass radii of the deuteron extracted from the fits of different $|t|$ ranges (see Fig. \ref{fig:deuteron-mass-radius-and-t-range}).
In Fig. \ref{fig:deuteron-mass-radius-and-t-range}, we see a relatively stronger dependence on the $|t|$-maximum boundary
of the fitting range, in comparison to the proton data.
One reason is probably that there is a stronger contribution from baryonic
or nuclear resonances for the deuteron data at the relatively higher photon energy.
By linearly extrapolating the fitting range to 0 GeV$^2$,
the mass radii of the deuteron at zero momentum transfer are obtained,
which are listed in Table \ref{tab:DeuteronMassRadius-CLAS}.
Finally, as suggested by the most near-threshold data of CLAS ($E_{\gamma}^{\rm aver.}=2$ GeV),
the mass radius of the deuteron is estimated to be $1.94\pm0.05$ fm.

LEPS collaboration also measured the near-threshold $\phi$ photoproduction on the deuteron \cite{Chang:2007fc}.
The statistical errors are much larger compared to the CLAS data.
The range of the variable $t$ is also much narrower,
which is not good for the extraction of the slope
of the differential cross section as a function of $|t|$.
Nevertheless, the $|t|$-dependence of LEPS data also can be described with the dipole form factor.
Fig. \ref{fig:diff-xsection-deuteron-target-LEPS} shows the fits compared to the cross-section data from LEPS.
The extracted mass radii of the deuteron at different energies are listed in Table \ref{tab:DeuteronMassRadius-LEPS}.
The LEPS data is closer to $t=0$ GeV$^2$, and the obtained deuteron mass radius
is consistent with the values extracted from CLAS data.

By looking at the average of all the near-threshold data of CLAS and LEPS,
the deuteron mass radius is combined to be $1.95 \pm 0.19$ fm.
Strikingly and similar to the proton case,
the mass radius of the deuteron is also smaller than the
world average of the charge radius of the deuteron
(CODATA-2010 average is $2.1424\pm0.0021$ fm).
The high precision data of the deuteron charge radius is from the Lamb shift
in the spectrum of the muonic deuterium,
which gives a value of $2.12562\pm0.00078$ fm \cite{CREMA:2016idx}.
For both the proton and the deuteron, the mass radii are about $0.1\sim 0.2$ fm
smaller than the charge radii.

In summary, first, we verified that the VMD model and low-energy QCD theorem
suggested by D. Kharzeev accurately describes the near-threshold quarkonium photoproduction.
We find that the VMD model and the dipole-form GFF reproduce well
the $|t|$-dependence of the differential cross section of $\phi$ photoproduction
on both the proton and the deuteron.
Nevertheless, there should be more studies on the GFF
and the mechanical properties of the hadronic system
encompassing more experimental data in the future.
Second, the mass radii of the proton and the deuteron are
determined from the parameterized GFF extracted from
the coherent and near-threshold $\phi$ photoproduction data from CLAS and LEPS collaborations.
The mass radii are found to be smaller than the charge radii for both the proton and the deuteron.
Third, the systematic uncertainty of the analysis is briefly investigated.
The mass radius extracted from the $|t|$-dependence of the cross section
is not sensitive to the size of the fitting range as long as $|t|<1$ GeV$^2$.
The dependence of the extracted mass radius on the photon energy is not big.
Judged by the CLAS data at several near-threshold energies,
the model uncertainty from not-at-threshold effects is
of the same order as the current statistical uncertainty
for both the proton and the deuteron.
Last, the systematic uncertainty related to the size of the vector meson
probe needs more future theoretical investigations. The near
threshold $\phi$-photoproduction data gives larger mass radius compared
to the J/$\psi$-photoproduction data \cite{Kharzeev:2021qkd}.

To further check the GFFs and the mass distributions of the proton and the nuclei,
the measurements of near-threshold photoproduction of J/$\psi$ and $\Upsilon$
will be performed at future Electron-Ion Colliders
\cite{AbdulKhalek:2021gbh,Accardi:2012qut,Chen:2018wyz,Chen:2020ijn,Anderle:2021wcy}.

\begin{acknowledgments}
This work is supported by the Strategic Priority Research Program of Chinese Academy of Sciences under the Grant Number XDB34030301,
the CAS Key Research Program of Frontier Sciences QYZDY-SSW-SLH006
and the National Natural Science Foundation of China under the Grant Numbers: 12005266, 11875296 and 11675223.
\end{acknowledgments}

\bibliographystyle{apsrev4-1}
\bibliography{refs}

\end{document}